\documentclass[sigconf]{acmart}
\AtBeginDocument{%
  }

\setcopyright{acmlicensed}
\copyrightyear{2025} 
\acmYear{2025} 
\setcopyright{acmlicensed}\acmConference[GeoIndustry '25]{The 4th ACM
 SIGSPATIAL International Workshop on Spatial Big Data and AI for Industrial
 Applications}{November 3--6, 2025}{Minneapolis, MN, USA}
 \acmBooktitle{The 4th ACM SIGSPATIAL International Workshop on Spatial Big
 Data and AI for Industrial Applications (GeoIndustry '25), November 3--6,
 2025, Minneapolis, MN, USA}
 \acmDOI{10.1145/3764919.3770875}
 \acmISBN{979-8-4007-2182-3/2025/11}

\usepackage{tabularx}

\begin{document}

\title{GBA-UBF : A Large-Scale and Fine-Grained Building Function Classification Dataset in the Greater Bay Area}
\renewcommand{\shorttitle}{GBA-UBF Dataset}

\author{
Chunsong Chen$^{\#}$,
Yichen Hou$^{\#}$,
Huan Chen$^{\#}$,
Junlin Li,
Rong Fu, \\
Qiushen Lai,
Yiping Chen$^{*}$,
Ting Han$^{*}$
}

\affiliation{
    \institution{School of Geospatial Engineering and Science, Sun Yat-sen University, Zhuhai, China}
    \city{}
    \country{}
}

\renewcommand{\shortauthors}{Chunsong Chen et al.}

\begin{abstract}
  Rapid urbanization in the Guangdong–Hong Kong–Macao Greater Bay Area (GBA) has created urgent demand for high-resolution, building-level functional data to support sustainable spatial planning. Existing land use datasets suffer from coarse granularity and difficulty in capturing intra-block heterogeneity. To this end, we present the \textbf{Greater Bay Area Urban Building Function Dataset (GBA-UBF)}, a large-scale, fine-grained dataset that assigns one of five functional categories to nearly four million buildings across six core GBA cities. We proposed a  \textbf{Multi-level Building Function Optimization (ML-BFO)} method by integrating Points of Interest (POI) records and building footprints through a three-stage pipeline: (1) candidate label generation using spatial overlay with proximity weighting, (2) iterative refinement based on neighborhood label autocorrelation, and (3) function-related correction informed by High-level POI buffers. To quantitatively validate results, we design the Building Function Matching Index (BFMI), which jointly measures categorical consistency and distributional similarity against POI-derived probability heatmaps. Comparative experiments demonstrate that GBA-UBF achieves significantly higher accuracy, with a BMFI of 0.58. This value markedly exceeds that of the baseline dataset and exhibits superior alignment with urban activity patterns. Field validation further confirms the dataset's semantic reliability and practical interpretability. The GBA-UBF dataset establishes a reproducible framework for building-level functional classification, bridging the gap between coarse land use maps and fine-grained urban analytics. The dataset is accessible at \textcolor{purple}{\href{https://github.com/chenchs0629/GBA\string-UBF}{https://github.com/chenchs0629/GBA-UBF}}, and the data will undergo continuous improvement and updates based on feedback from the community.
\end{abstract}


\begin{teaserfigure}
  \centering
  \includegraphics[width=0.87\textwidth]{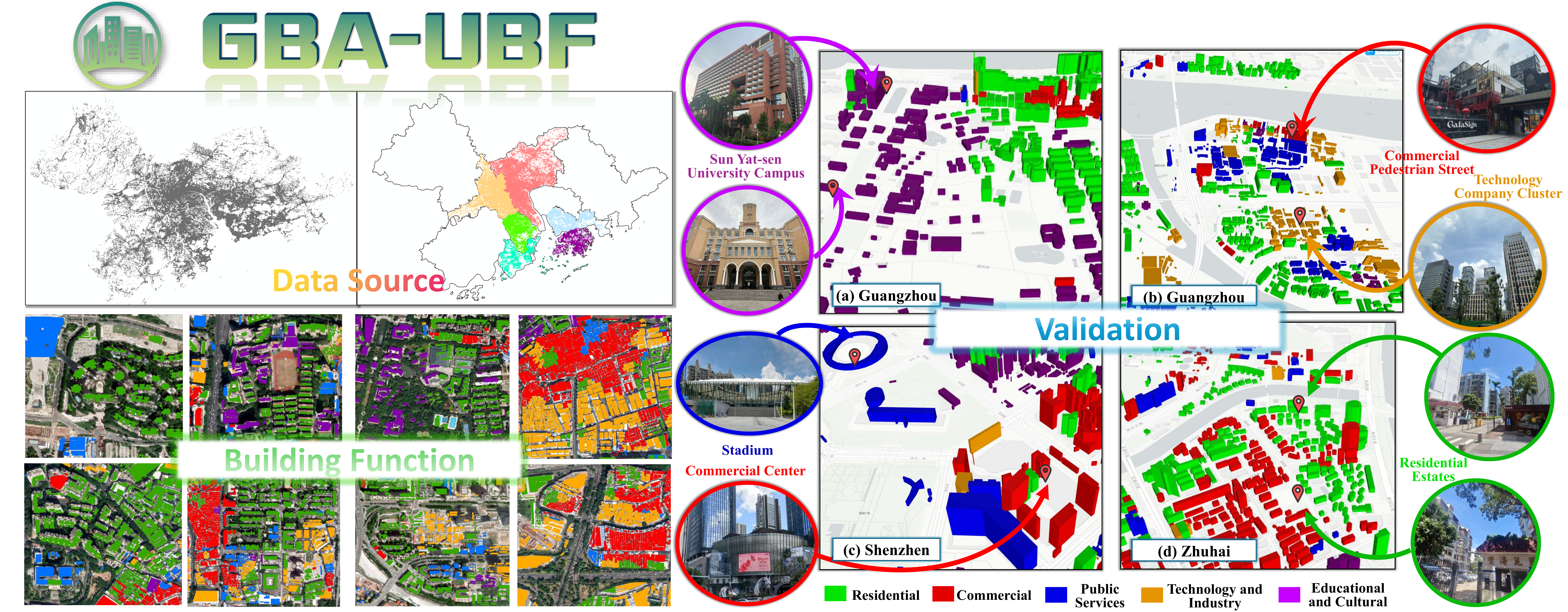}
  \setlength{\abovecaptionskip}{0.1cm}
  \setlength{\belowcaptionskip}{-0.1cm}
  \caption{Overview of GBA-UBF Dataset, a building-level function dataset for the Greater Bay Area. GBA-UBF provides city-wide, building-scale labels that overcome the coarse granularity of parcel maps. It covers ~4 million buildings across six core GBA cities with five unified classes: Residential, Commercial, Public Services, Technology \& Industry, Educational \& Cultural.}
  \label{fig:overview}
  \vspace{0.3cm}
\end{teaserfigure}


\maketitle

\section{Introduction}

Accurate classification of urban land use and building functions plays a crucial role in spatial planning and sustainable development. Fine-grained functional mapping not only supports the optimization of urban spatial structures but also guides infrastructure deployment and enables evidence-based governance. 

In recent years, the proliferation of social media and mobile internet has made user-generated content (UGC) an important geospatial data source. With its near real-time updates, broad coverage, and strong linkages to human activities, UGC complements traditional Earth observation (EO) datasets, such as high-resolution satellite imagery. For land-use inference, UGC enriches socio-economic semantics that are often missing in conventional EO products and supports refined functional zoning \cite{tu2021user}. Prior studies that integrate UGC (e.g., Points of Interest (POI), GPS trajectories, and OpenStreetMap) with EO imagery have reported notable advances in urban land-use classification and functional mapping \cite{chen2021mapping,li2025enhanced,xiong2025mapping}.

Nevertheless, mainstream land-use datasets remain parcel-based, where blocks are annotated with functions such as residential, commercial, industrial, public services, or green spaces. While parcel-level maps underpin critical planning tasks such as the delineation of Urban Functional Zones (UFZs) \cite{hersperger2018urban} and assessments of urban growth dynamics \cite{puertas2014assessing}, they suffer from several limitations:
\begin{enumerate}
    \item \textbf{Inconsistent classification standards.} heterogeneous taxonomies hinder benchmarking and validation.
    \item \textbf{Inconsistent parcel delineation.} variations in road-based segmentation rules lead to spatial inconsistencies and scale effects \cite{hecht2015automatic}.
    \item \textbf{Multi-functionality and scale mismatch.} parcels often contain hundreds of buildings with mixed uses, undermining assumptions of homogeneity.
\end{enumerate}
These limitations constrain the robustness and applicability of parcel-level data for fine-scale urban research.

Buildings, as the atomic units of cities, accommodate living, working, commerce, education, and recreation, thereby encoding fine-grained land use information \cite{zhuo2019identifying}. Studies leveraging POI, remote sensing imagery, and UGC have demonstrated the potential of building-level functional inference \cite{lin2021identifying,wang2021building}, enabling applications ranging from emergency response \cite{pourghasemi2020assessment} and policy-making \cite{zhong2014inferring} to resource management \cite{srivastava2018multilabel}.

Recent advances in spatiotemporal big data (e.g., Google Maps, Amap, and Baidu Maps) have also made large-scale building-level analysis feasible. POI datasets provide semantic attributes, while building footprints capture precise geometries, together offering unprecedented opportunities to move beyond coarse parcel-based datasets toward detailed, reproducible urban function mapping. However, existing research often remains limited to small-scale case studies, lacks standardized classification schemes, and rarely validates across cities with diverse morphologies.

Against this backdrop, the Guangdong–Hong Kong–Macao Greater Bay Area (GBA) emerges as a compelling context to address these research gaps. China released the Outline Development Plan for the GBA in 2019, which set forth the strategic goal of developing the GBA into a world-class bay area. Among the world’s four major bay areas, the GBA is the largest in land area, population, and industrial agglomeration, thereby facing highly complex challenges in urban planning, demographic dynamics, human activities, and community infrastructure \cite{hui2020deciphering}. Given the GBA's extreme population density, diverse economic activities, and fragmented urban expansion, traditional land-cover mapping alone is insufficient. Consequently, building-level functional classification is essential for formulating responsive policies and adaptive urban management. Meanwhile, the GBA comprises 11 cities or special administrative regions with distinct urban forms and industries, making an urban building function dataset built upon this urban agglomeration both representative and unique.

\begin{figure}[!t]
    \setlength{\abovecaptionskip}{0.1cm}
    \setlength{\belowcaptionskip}{-0.1cm}
    \centering
    \includegraphics[width=0.48\textwidth]{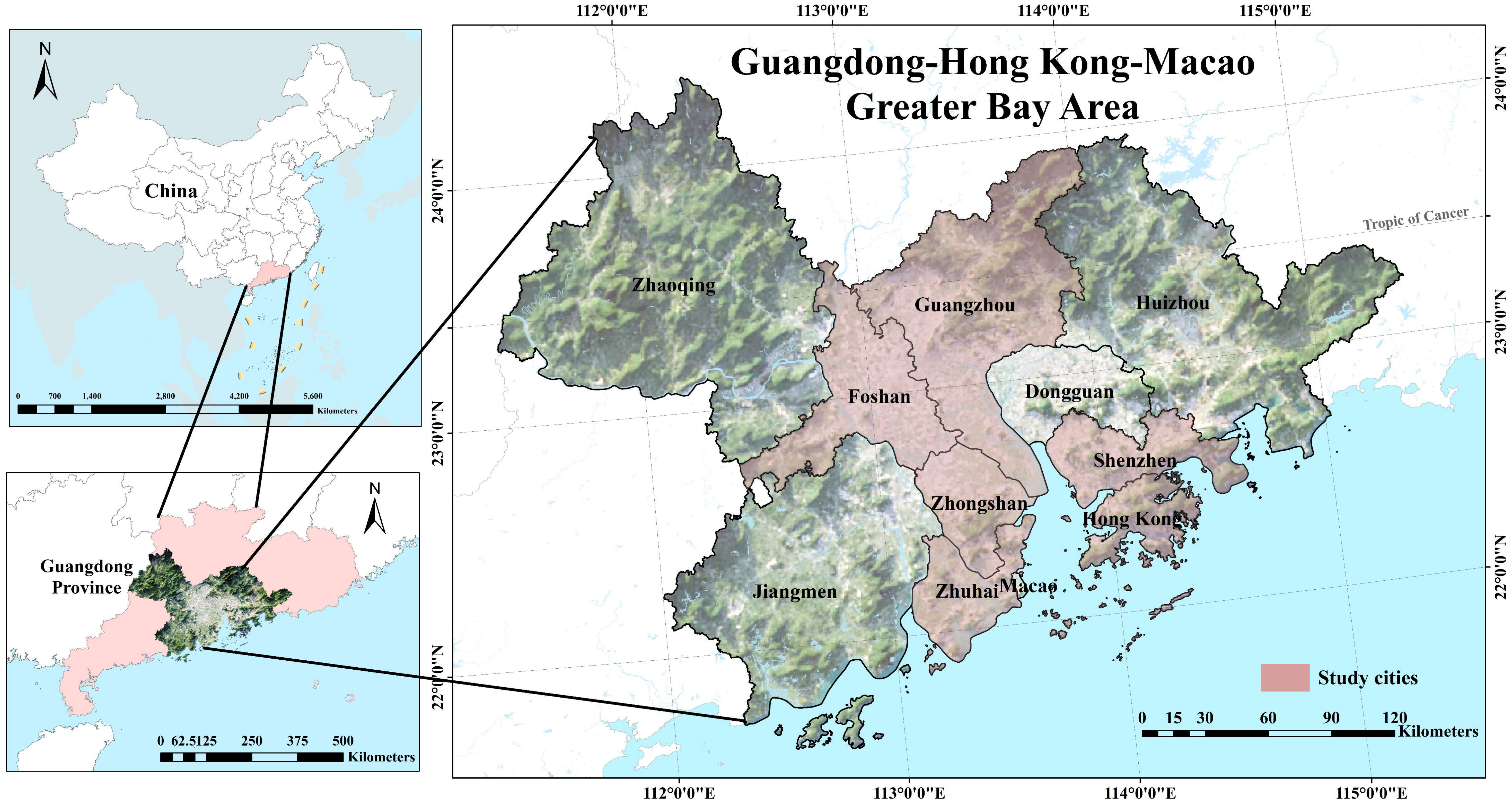} 
    \caption{Geographical Location of the GBA and Study Area.}
    \label{fig:SA}
\end{figure}

However, there remains no large-scale, fine-grained building function dataset for the GBA. To fill this gap, this study constructs the new \textbf{Greater Bay Area Urabn Building Function Dataset (GBA-UBF)}, which covers nearly four million buildings across six core cities. The dataset is generated through a reproducible three-stage pipeline, which we have named as \textbf{Multi-level Building Function Optimization (ML-BFO)} method: (1) Candidate label generation via spatial overlay of POI and building footprints with proximity-weighted scoring. (2) Iterative refinement leveraging neighborhood label autocorrelation to enhance spatial consistency. (3) Function-related correction incorporating advanced POI buffers to capture the influence of landmark facilities.

We further propose the Building Function Matching Index (BFMI), a novel evaluation metric that integrates categorical consistency with distributional similarity against POI probability heatmaps. Comparative analysis with the EULUC-China 2.0 dataset demonstrates the superior granularity and accuracy of GBA-UBF, particularly in high-density and mixed-use areas. Field validation confirms its semantic reliability and practical interpretability.

\begin{figure*}[!t]
    \setlength{\abovecaptionskip}{0.1cm}
    \setlength{\belowcaptionskip}{-0.1cm}
    \centering
    \includegraphics[width=0.88\textwidth]{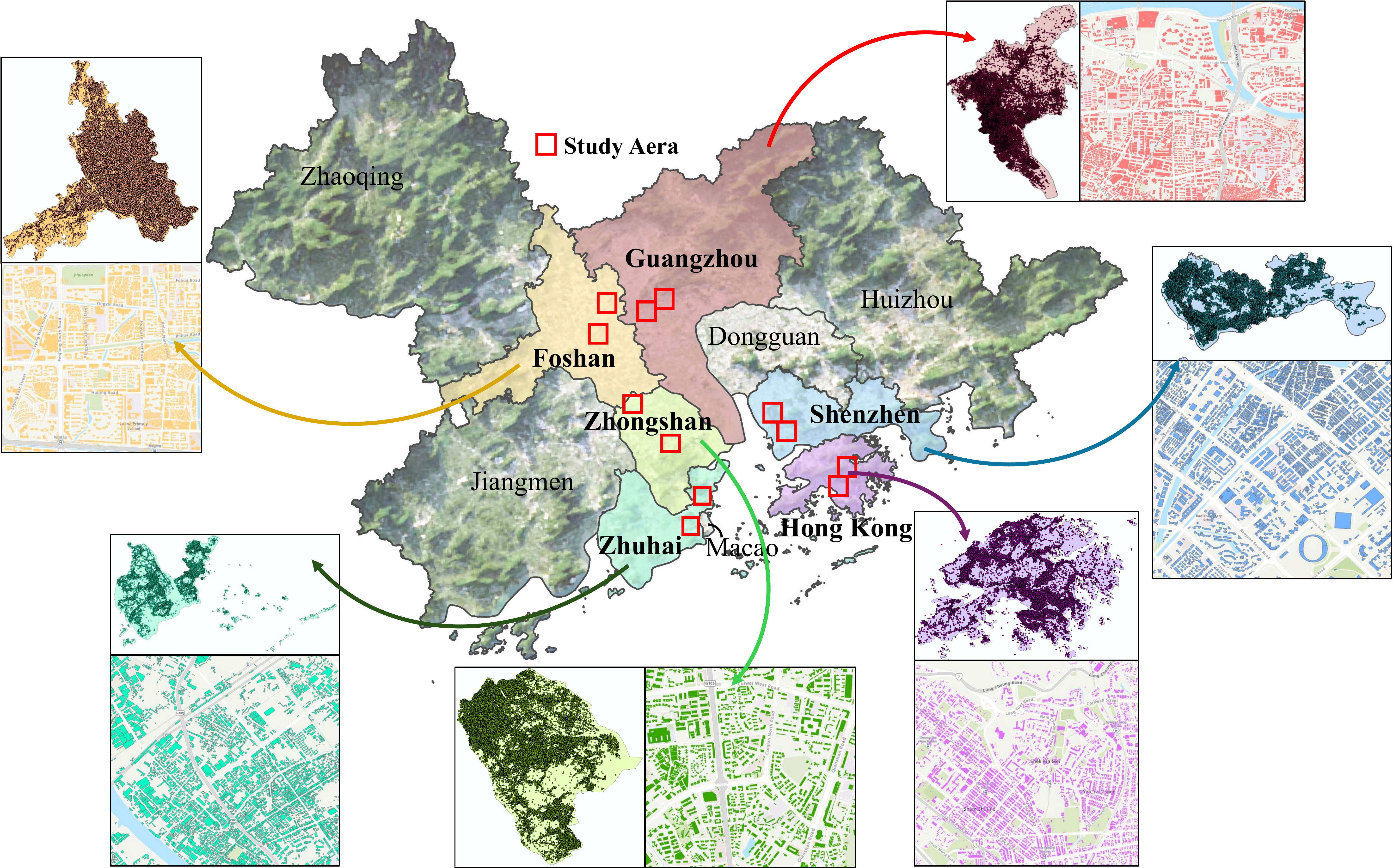}
    \caption{Location and Data Overview of the Study Area within the Greater Bay Area. Different colors represent different cities, and for each city we show both the distribution of POI and representative building footprints.}
    \label{fig:location}
\end{figure*}
\section{Related works}

\subsection{Urban Land Use Dataset}  

Land use and land cover (LULC) mapping is foundational to urban and regional planning and is now a core enabler of smart, sustainable cities \cite{GeoAI_Planning2024}. Recent advances in very-high-resolution (VHR) Earth observation, together with the rapid expansion and cross-domain integration of multimodal geospatial data, have produced a new generation of LULC products that are spatially explicit, temporally consistent, and operational at scale. The representative datasets include GlobalLand30 \cite{chen2015global}, FROM-GLC30 and FROM-GLC10 \cite{gong2013finer,gong2019stable}, the ESA WorldCover (10 m) suite \cite{zanaga2022esa}, GLC-FCS10 (10 m) derived from Sentinel imagery on Google Earth Engine \cite{zhang2025glc_fcs10}, 3-m national maps of China generated using public 10-m land cover maps as training data \cite{dong2020improving}, and the SinoLC-1 (1 m) product \cite{li2023sinolc}. These products exemplify the field’s shift toward higher spatial resolution and more frequent updates in global LULC mapping.

In contrast, urban land use (ULU) is substantially harder to delineate than land cover due to semantic ambiguity, temporal volatility, and pervasive mixed use. The intensification of human-driven land transformation makes accurate, intra-urban ULU characterization scientifically important and operationally urgent. High-resolution Earth observation provides the spatial detail and radiometric quality required for fine-scale ULU mapping \cite{han2025cityinsight}. Over the past decade, methodological advances have systematically exploited textural, geometric, and contextual cues in these data, yielding increasingly discriminative representations of urban form and function, from object-based image analysis to deep context modeling \cite{bian2017fusing,shi2021object,cuypers2023land}.

However, spectral-only classification quickly reaches the information ceiling of imagery and cannot deliver the granularity demanded by complex urban scenes. The proliferation of spatiotemporal big data and social sensing has enriched land-use representations and enabled more refined mapping. OpenStreetMap (OSM) is widely used to regularize labels and improve classification \cite{fonte2017assessing,yang2017open}. However, its inherent incompleteness and positional uncertainty remain non-trivial constraints \cite{zhou2022exploring}. UGC enables crowdsourced, continuously updated, activity-aware geospatial datasets \cite{wu2021comprehensive}. In particular, points of interest (POI) supply semantic labels largely absent from Earth-observation imagery, providing supervision/weak supervision and reference annotations for urban land-use classification \cite{wu2021identifying,liu2020investigating}. Fusing POI-derived features with VHR imagery and OSM consistently yields more accurate and more robust land use maps \cite{andrade2020poi,liu2020recognizing,xiong2025mapping}.

\subsection{Urban Building Function Analysis}

Urban Functional Zoning (UFZ) refers to the division of urban areas into zones according to their dominant functions, such as residential, commercial, industrial, and public service areas. With the acceleration of global urbanization, the accurate identification of UFZs is critical for optimizing urban spatial structure, guiding planning, and improving resource allocation \cite{liu2021classification}. Traditional functional zoning has relied primarily on land-use planning maps and census data, which suffer from long update cycles and coarse resolution, limiting their ability to capture rapidly changing urban morphologies.

With the widespread adoption of Geographic Information Systems, remote sensing, and location-based services, researchers have increasingly turned to spatial big data for UFZ identification. Among these, POI provide rich semantic features that offer fine-grained descriptions of urban functions, making POI-based methods a major research focus. Many studies have employed POI spatial distributions and categorical attributes, often combined with kernel density estimation, clustering, or other statistical methods, to delineate UFZs and reveal their spatial patterns \cite{luo2023urban}.

For instance, Zhang et al. \cite{zhang2018identifying} automatically delineated the functional zones of Hangzhou using POI data, while Miao et al. \cite{miao2021analyzing} extracted urban spatial patterns from Sina Weibo POI. Hong et al. \cite{hong2019hierarchical} highlighted the role of roads in shaping functional distributions by integrating POI data with OpenStreetMap. Beyond single-source approaches, the integration of multi-source data has further advanced UFZ research. Zhang et al. \cite{zhang2017hierarchical} proposed a hierarchical semantic framework that fused very-high-resolution (VHR) imagery with POI, while Liu et al. \cite{liu2020recognizing} combined Landsat imagery with taxi GPS and POI to recognize UFZs. These studies collectively demonstrate the potential of fusing multi-source spatial and semantic data to improve the accuracy and interpretability of UFZ identification across different urban scales and contexts.

\section{Dataset and Study Area}

\subsection{Study Area}

The Guangdong–Hong Kong–Macao Greater Bay Area (GBA), anchored in the Pearl River Delta, spans 56,000 km² and is home to over 86 million residents, making it one of China’s most urbanized and densely populated regions. The GBA includes nine Guangdong prefectures: Guangzhou, Shenzhen, Zhuhai, Foshan, Zhongshan, Dongguan, Huizhou, Jiangmen, Zhaoqing—as well as the Hong Kong and Macao Special Administrative Regions (SARs).

In terms of functional specialization, Shenzhen is positioned as the innovation hub, Guangzhou as the integrated services and transport gateway, Hong Kong as the global financial and shipping center, Macao as a world tourism and leisure hub, and the Dongguan and Foshan area as a center for advanced manufacturing. With seamless transport connectivity envisioned to link these key nodes into a high-productivity megalopolis, the GBA faces significant governance challenges in coordinating a nearly 100 million strong, polycentric region with overlapping jurisdictions. The geographic extent of the GBA is shown in Fig.~\ref{fig:SA}.

\begin{table}[!t]
\setlength{\abovecaptionskip}{0.1cm}
\setlength{\belowcaptionskip}{-0.1cm}
\centering
\caption{Comparative Statistics of POI Quantity and Building Footprints in Selected Cities' study areas.}
\label{tab:city_data}
\begin{tabular}{ccc}
\toprule
\textbf{City} & \textbf{POI Quantity} & \textbf{Building Footprints} \\
\midrule
\textbf{Guangzhou}   & 1,499,128 & 1,351,299 \\
\textbf{Shenzhen}    & 1,413,510 &   741,612 \\
\textbf{Zhongshan}   &   404,371 &   391,665 \\
\textbf{Zhuhai}      &   183,090 &   171,031 \\
\textbf{Foshan}      &   771,242 & 1,038,253 \\
\textbf{Hong Kong}   &   546,720 &   175,905 \\
\bottomrule
\end{tabular}
\end{table}

\begin{table}[!t]
\setlength{\abovecaptionskip}{0.1cm}
\setlength{\belowcaptionskip}{-0.3cm}
\centering
\caption{Correspondence between Building Function Types and Baidu POI Classification.}
\label{tab:function-classification}
\footnotesize
\begin{tabularx}{\columnwidth}{>{\bfseries}p{2.8cm}X}
\toprule
\textbf{Building Function Type} & \textbf{Baidu POI Primary Classification System} \\
\midrule
Residential & Real Estate \\
Commercial & Food, Hotel, Shopping, Life Services, Beauty Services, Recreation Center, Sports and Fitness Center, Car Services, Financial Industry \\
Public Services & Life Services, Culture and Media, Government Agencies, Transportation, Healthcare, Tourist Attractions \\
Technology and Industry & Companies and Enterprises \\
Educational and Cultural & Education and Training Venue\\
\bottomrule
\end{tabularx}
\end{table}

Urbanization in the GBA is well advanced, but ongoing integration is blurring inter-city boundaries and accelerating the urbanization of peri-core areas. Shenzhen (density >7,000 inhab/km²) and Guangzhou (density >2,000 inhab/km²) exemplify extreme urban density and intensifying mixed land use \cite{qiao2024assessment}. The functional distinctions between municipalities are becoming less clear, making traditional administrative zoning inadequate for coordinated governance and efficient resource allocation. Additionally, the varying development stages and functional mandates across cities lead to diverse spatial distributions of residential, commercial, industrial, technological, public service, and educational facilities.

To address this complexity, the study focuses on the central urban areas of six cities—Guangzhou, Shenzhen, Hong Kong, Zhuhai, Zhongshan, and Foshan—chosen for data availability, analytical feasibility, and structural comparability. Guangzhou and Shenzhen represent the traditional urban core and innovation-driven metropolis, respectively, while Hong Kong exemplifies compact vertical urbanism. Foshan and Zhongshan are characterized by manufacturing-focused zoning, and Zhuhai offers a tourism-oriented, livable morphology. These cities span various urban tiers, ensuring the dataset’s representativeness and cross-city comparability. The study area location within the GBA and data overview are presented in Fig.~\ref{fig:location}.

\subsection{Data Sources}

To support urban functional analysis, we integrate two complementary data sources:
\begin{itemize}
  \item \textbf{Building Footprints:} Aggregated from Baidu Maps\footnote{\url{https://map.baidu.com/}} and OSM\footnote{\url{https://www.openstreetmap.org/}}. We unify the data formats, remove duplicate polygons, and reproject all geometries, yielding accurate building footprints for further analysis.
  \item \textbf{POI Data:} POIs are sourced from Baidu Maps. Each POI provides a georeferenced location and rich attributes (name, address, category), enabling inference of urban building functions. As curated records of real-world entities on a high-traffic commercial platform, Baidu POI serve as timely, activity-aware proxies for land use in China. For functional labeling, we parse the Baidu industrial‐type field into primary and secondary categories and harmonize it to a 19-class primary taxonomy adopted in this paper.
\end{itemize}
The quantity of POI and building footprints for each city's study areas is summarized in Tab.~\ref{tab:city_data}.

\begin{figure*}[!t]
    \setlength{\abovecaptionskip}{0.1cm}
    \setlength{\belowcaptionskip}{-0.2cm}
   \centering
   \includegraphics[width=0.85\textwidth]{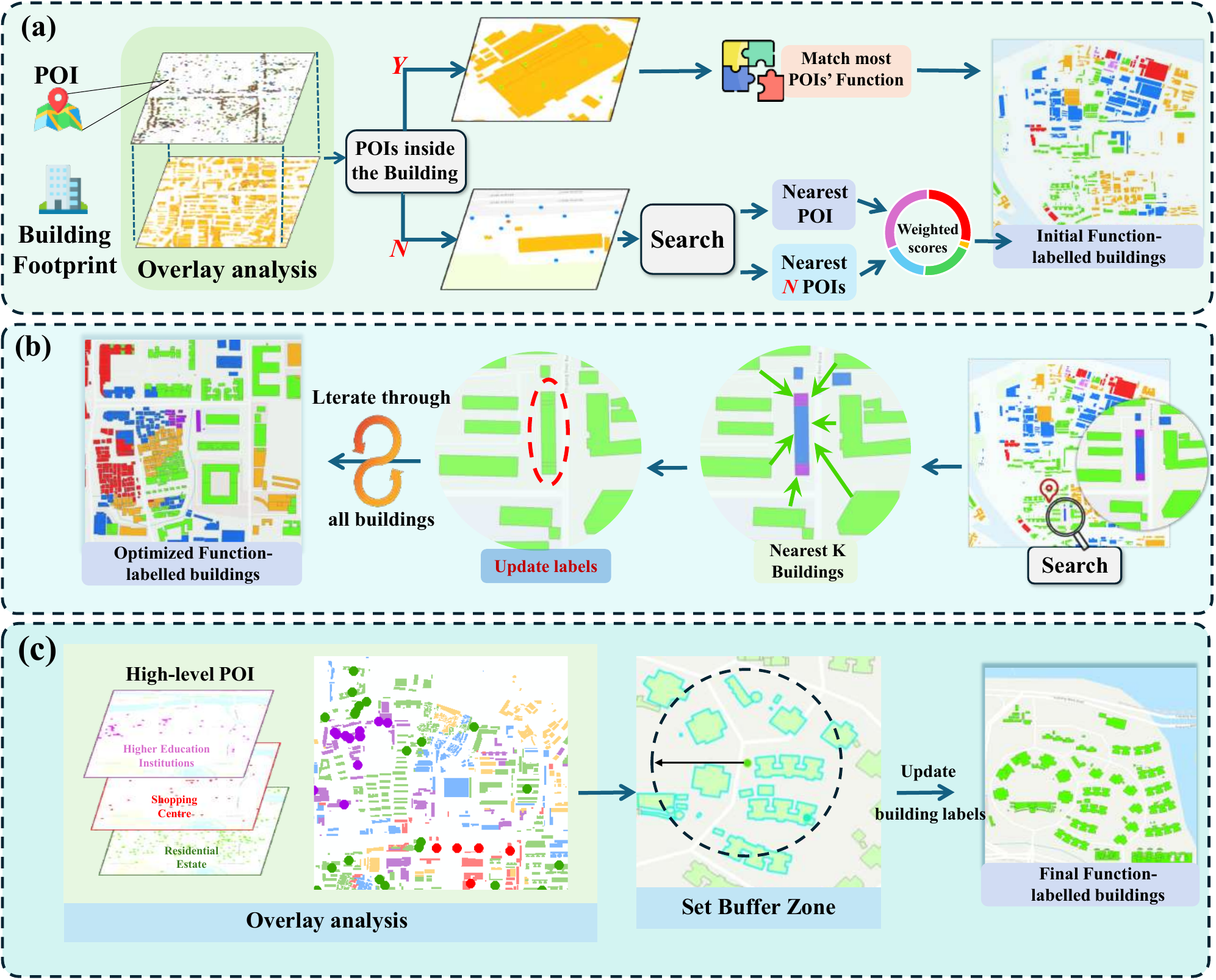} 
    \caption{Overall of dataset generation framework. We use a reproducible 3-stage pipeline: (a) Candidate Function Label Generation; (b) Iterative Label Distribution Refinement; and (c) Function Related Label Correction.}
   \label{fig: method}
\end{figure*}

\subsection{POI Reclassification with Building Function}

Baidu POI data consist of 19 primary categories\footnote{dining, hotels, shopping, life services, beauty services, tourist attractions, leisure and entertainment, sports and fitness, education and training, culture and media, medical care, cycling services, transportation facilities, finance, real estate, companies and enterprises, government agencies, entrances/exits, and natural features}. For the purpose of building function identification, we manually exclude three non-building categories: entrances/exits, roads, and natural features. To improve the accuracy of delineating urban spatial structure, and following the National Standard for Urban Land Use and Planning (GB 50137-2011), we further remove land-use types that lack physical buildings and consolidate the remaining categories. As a result, all urban buildings are classified into five functional types: Commercial Services, Residential, Technology and Industry, Educational and Cultural, and Public Services.

Tab.~\ref{tab:function-classification} lists each functional type and its corresponding POI sub-categories. Specifically, Commercial Services encompass daily commercial premises supporting business and consumption; Public Services include government, healthcare, transportation, cultural, and public facilities; Residential refers to dwelling-oriented areas; Technology and Industry cover science parks, industrial estates, factories, and office complexes; and Educational and Cultural represent schools, universities, and research institutes.

This taxonomy is grounded in theories of urban land use and functional zoning, integrating POI attributes with spatial distribution patterns to capture the dominant socio-economic role of each building. In practice, buildings often exhibit mixed functions, such as retail stores embedded in residential compounds. However, this dataset does not include a mixed-use category, as the complexity of such classification greatly increases annotation difficulty, and scholarly consensus on its definition remains limited \cite{evans2019modelling}. Adopting a dominant-function strategy ensures classification stability, reproducibility, and comparability across cities.
\section{Methodology}

In this study, we proposed a \textbf{Multi-level Building Function Optimization (ML-BFO)} method. It consists of three core stages: (1) \textbf{Candidate function label generation}; (2) \textbf{Iterative label distribution refinement}; and (3) \textbf{Function-related label correction}, as shown in Fig.~\ref{fig: method}. Moreover, we further design the \textbf{Building Function Matching Index (BFMI)} to quantitatively assess the reliability of the dataset. The details are as follows.

\subsection{Candidate Function Label Generation}




In large-scale urban spatial analysis, accurately identifying and labeling the functional attributes of buildings is essential for understanding urban structure and dynamics. Due to the high density and large volume of POI data provided by Baidu Maps, which far exceed the number of building footprints, we propose a candidate function label generation method via spatial overlay representation.  

Let a building $B_i$ contain the POI set $\mathcal{P}_i = \{p_{i1}, p_{i2}, \dots, p_{in}\}$, where each POI $p_{ij}$ belongs to a category $c(p_{ij}) \in \mathcal{C}, \mathcal{C} = \{1,2,3,4,5\}$.  

\subsubsection*{\textbf{Spatial proximity score}}
To quantify the spatial relevance between a POI and its nearby building, we define a \textit{spatial proximity score} as:
\begin{equation}
    \alpha_{ij} = \frac{1}{1 + d(p_{ij}, B_i)},
\end{equation}
where $d(p_{ij}, B_i)$ denotes the shortest Euclidean distance between POI $p_{ij}$ and the boundary of building $B_i$.  
A smaller distance implies that the POI is more closely associated with the building—either functionally or spatially—indicating a higher likelihood that the POI’s category reflects the building’s actual use. Conversely, distant POIs have weaker spatial linkage and thus contribute less to the functional label.

\subsubsection*{\textbf{Category-specific weight}}
Since the frequency of different POI categories varies across study aera, the importance of each category is further modulated by a \textit{category-specific weight} $w_c(B)$, which reflects the relative representativeness of category $c$ within the local block $B$.  
For each block $B$, we compute the category count $\text{freq}_c(B)$ and the total number of POIs $N(B)$.  
The weight is defined with a power decay function as:
\begin{equation}
    w_c(B) = \frac{1}{\left( \text{freq}_c(B) + 1 \right)^{\alpha}}, \quad \alpha \in [0.3, 1.0].
\end{equation}
This formulation reduces the dominance of highly frequent categories while preserving the contribution of rare but strongly indicative POIs. The parameter $\alpha$ is manually set based on different study areas.

\subsubsection*{\textbf{Composite score and label assignment}}
The composite score of building $B_i$ for category $c$ is computed as:
\begin{equation}
    S_i(c) = \sum_{p_{ij} \in \mathcal{P}_i, \, c(p_{ij})=c} w_c(B) \cdot \alpha_{ij}.
\end{equation}
Finally, the dominant functional label is assigned according to the maximum score:
\begin{equation}
    L_i = \arg\max_{c \in \mathcal{C}} S_i(c).
\end{equation}
This method generates stable candidate labels and provides a robust basis for subsequent refinement.


\subsection{Iterative Label Distribution Refinement}

The initial POI-derived building labels may be noisy, especially in mixed-use or data-sparse areas. To improve label accuracy and robustness, we introduce an iterative refinement strategy based on spatial autocorrelation, leveraging the assumption that nearby buildings tend to share similar functions.

Formally, let the building set be $B = \{b_1, b_2, \dots, b_n\}$ with initial labels $L^{(0)}(b_i)$. For each building $b_i$, denote its $k$ nearest neighbors as $N_k(b_i)$. At iteration $t$, the label update rule is defined as:
\begin{equation}
L_i^{(t+1)} =
\begin{cases}
\arg\max_{l \in \mathcal{L}} \sum_{j \in N_k(i)} \mathbf{1}(L_j^{(t)} = l), & \text{if } \Phi_i^{(t)} > \tfrac{k}{2}, \\[1ex]
L_i^{(t)}, & \text{otherwise.}
\end{cases}
\end{equation}
where $\Phi_i^{(t)} = \max_{l \in \mathcal{L}} \sum_{j \in N_k(i)} \mathbf{1}(L_j^{(t)} = l)$ denotes the maximum count of label $l$ among the $k$ neighbors of building $b_i$.

This process repeats until convergence (no significant label changes) or until a preset maximum number of iterations is reached. By integrating neighborhood label distributions, the method enhances local consistency, reinforces spatial clustering, and significantly improves the precision of urban functional zoning.


\subsection{Function Related Label Correction}

To further enhance label accuracy, we propose a correction mechanism that integrates High-level POI and their surrounding buffer zones. High-level POI are defined as landmark facilities with strong functional implications, such as large commercial centers, residential complexes, and higher-education institutions. These facilities not only determine the function of the buildings they occupy but also exert influence on the surrounding urban fabric. To capture this effect, we generate buffers of varying radii for each High-level POI, representing its spatial sphere of influence.

During the correction phase, each building with an initial label is examined against these High-level POI buffers. If a building falls within the buffer of a given POI, its label is updated to match the category of POI, reflecting its functional association. For instance, a building initially labeled as "Residential" may be reclassified as "Commercial" if it lies within the buffer of a large shopping mall. By leveraging the radiating effects of landmark facilities, this method introduces higher-level semantic correction, improving both the accuracy and granularity of functional zoning.


\subsection{Evaluation Metric System Design}

To assess the accuracy of the derived urban building function classifications, we introduce the \textbf{Building Function Matching Index (BFMI)}. Considering the lack of reliable ground-truth data and the fact that existing land-use datasets cannot capture the fine-grained characteristics of building-level functional patterns, traditional evaluation metrics (e.g., precision, recall, F1-score) are not directly applicable. Therefore, BFMI provides a practical alternative for quantitative evaluation. It quantitatively measures the alignment between building-level classification outcomes and POI-derived spatial probability distributions.


\subsubsection{\textbf{POI Probability Heat Map Generation.}} 

Specifically, POI records corresponding to the five functional categories are processed using Kernel Density Estimation (KDE) to generate class-specific heat maps. These heat maps serve as reference standards, capturing the spatial clustering and distributional characteristics of each functional type. For each functional class $c \in \{1, \dots, 5\}$, KDE is applied to the raw POI to obtain the density function $D_c(x)$. Then, we apply a softmax transformation to derive normalized probability surfaces:  
\begin{equation}
P_{\text{poi},c}(x) = \frac{D_c(x)}{\sum_{k=1}^{5} D_k(x) + \epsilon},
\end{equation}
where $\epsilon$ is a smoothing term.  

\subsubsection{\textbf{Reference Label Approximation.}}

For each building polygon $B_i$, we extract pixel values from the POI-based probability raster within its footprint and compute zonal statistics. By averaging across the polygon, we obtain a probability distribution vector over the five functional classes:
\begin{equation}
    \bar{P}_{\text{poi}}(B_i) = (\bar{p}_1, \bar{p}_2, \dots, \bar{p}_5), \quad \sum_{c=1}^{5} \bar{p}_c = 1,
\end{equation}
where $\bar{p}_c$ represents the average probability that building $B_i$ belongs to functional class $c$, reflecting the predictive tendency derived from the POI raster.

Meanwhile, the approximation label $y_i$ of building $B_i$ is encoded as a one-hot vector:
\begin{equation}
    P_{\text{bld}}(B_i) = \text{one\_hot}(y_i).
\end{equation}
For example, if building $B_i$ is Residential, the corresponding one-hot vector is $(0, 1, 0, 0, 0)$. This encoding enables direct vector space comparison between the true building label and the POI-derived probability distribution.

\begin{figure}[!t]
    \setlength{\abovecaptionskip}{0.1cm}
    \setlength{\belowcaptionskip}{-0.2cm}
    \centering
    \includegraphics[width=0.48\textwidth]{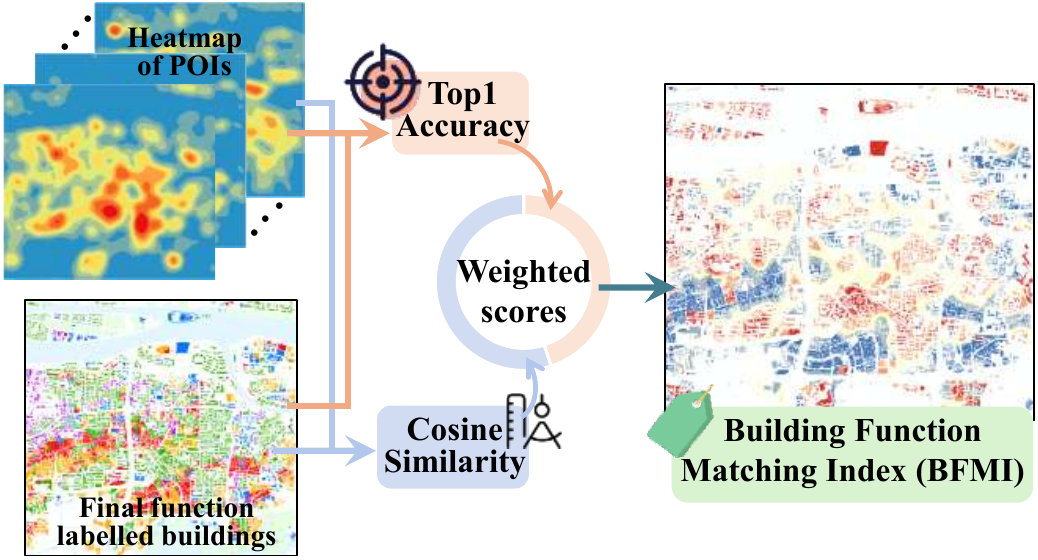}
    \caption{Construction of the Building Function Matching Index (BFMI). POI records are converted to class-specific KDE heatmaps and compared with the final building-level labels. For each building we compute (1) Top-1 Consistency with the dominant POI class and (2) Cosine Similarity between the one-hot label and the POI-derived probability vector, to generate the BFMI score map.}
    \label{fig:4.4}
\end{figure}

\subsubsection{\textbf{Evaluation Metrics.}}  

Ours BFMI integrates categorical accuracy and distributional consistency, as shown in Fig.~\ref{fig:4.4}. For each building $B_i$, two components are computed:

\paragraph{\textbf{(a) Top-1 Consistency}.} This metric checks whether the predicted functional class matches the reference label:
\begin{equation}
\text{Top1}(B_i) = \mathbf{1}\{\arg \max_c \hat{p}_c = y_i\},
\end{equation}
where $\hat{p}_c$ denotes the predicted probability for class $c$, and $y_i$ is the reference class of building $B_i$.

\paragraph{\textbf{(b) Distributional Similarity}.} To measure alignment between the predicted probabilities and the POI-derived reference distribution, we adopt cosine similarity:
\begin{equation}
\text{Cos}(B_i) = 
\frac{\bar{P}_{\text{poi}} \cdot P_{\text{bld}}}
     {\|\bar{P}_{\text{poi}}\| \; \|P_{\text{bld}}\|}
= \frac{\bar{p}_{y_i}}{\sqrt{\sum_{c=1}^{C} \bar{p}_c^2}},
\end{equation}
where $\bar{P}_{\text{poi}}$ is the POI-based probability vector and $P_{\text{bld}}$ is the one-hot encoding of the predicted label. $\text{Cos}(B_i)$ ranges from 0 to 1, with higher values indicating stronger agreement between the prediction and the dominant POI distribution.

\paragraph{\textbf{(c) Comprehensive Index (BFMI)}} Finally, the two components are fused into the BFMI:
\begin{equation}
\text{BFMI} = \frac{1}{N} \sum_{i=1}^N \Big[w \cdot \text{Top1}(B_i) + (1-w) \cdot \text{Cos}(B_i)\Big],
\end{equation}
where $w \in [0,1]$ is a balancing weight.In this study we set $w=0.5$ as default, giving equal importance to Top-1 consistency and cosine similarity.

\section{Results and Discussion}

\subsection{Dataset Description}

Our Greater Bay Area Urban Building Function (GBA-UBF) Dataset delineates five major building functions: \textbf{\textit{Commercial Services}, \textit{Residential}, \textit{Public Services}, \textit{Technology and Industry}, and \textit{Educational and Cultural}}, as shown in Tab.~\ref{tab:function-classification}. Each building footprint is explicitly assigned to one functional category, ensuring a consistent and fine-grained representation of the urban built environment. Data compilation required half a year of effort by a five-member team, involving POI collection, footprint cleaning, algorithm refinement, and manual validation. To standardize processing, the dataset is first organized into five shapefiles and then merged into a unified geospatial layer with class attributes. Building footprints are stored at the polygon level, supporting precise delineation and detailed analyzes of function allocation and validation. The dataset also incorporates building-height attributes, enabling statistical assessment and spatial modeling.

\begin{figure}[!t]
    \setlength{\abovecaptionskip}{0.1cm}
    \setlength{\belowcaptionskip}{-0.4cm}
    \centering
    \includegraphics[width=0.95\columnwidth]{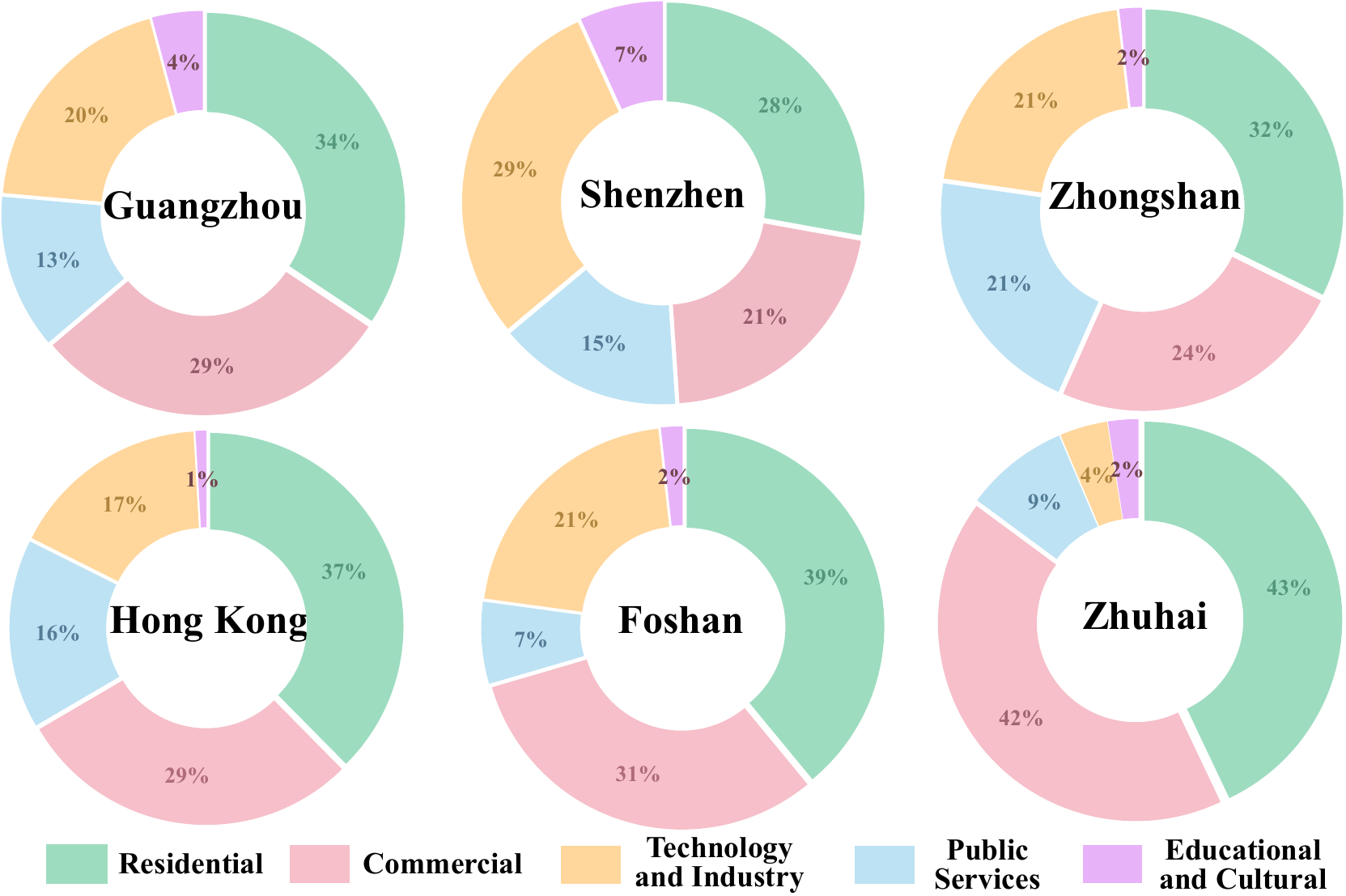}
    \caption{Distribution of Building Functional Categories in the Central Urban Areas of Six GBA Cities.}
    \label{fig:pro}
\end{figure}

In terms of file structure, each record contains the building geometry (polygon), functional class, spatial reference system, and exact geographic coordinates of vertices, ensuring compatibility with standard GIS workflows. The distribution of building function categories across the six cities is illustrated in Fig.~\ref{fig:pro}. The dataset covers the central urban cores of six representative GBA cities: Guangzhou, Shenzhen, Hong Kong, Foshan, Zhongshan, and Zhuhai. For each city, two representative zones of approximately $5 \times 5$ km are selected for dataset construction. The processed areas include 45,000 buildings in Guangzhou, 38,000 in Shenzhen, 20,000 in Hong Kong, 69,000 in Foshan, 21,000 in Zhongshan, and 13,000 in Zhuhai, amounting to more than 300 km$^2$. 

\begin{figure*}[!t]
    \setlength{\abovecaptionskip}{0.1cm}
    \setlength{\belowcaptionskip}{-0.3cm}
    \centering
    \includegraphics[width=0.8\textwidth]{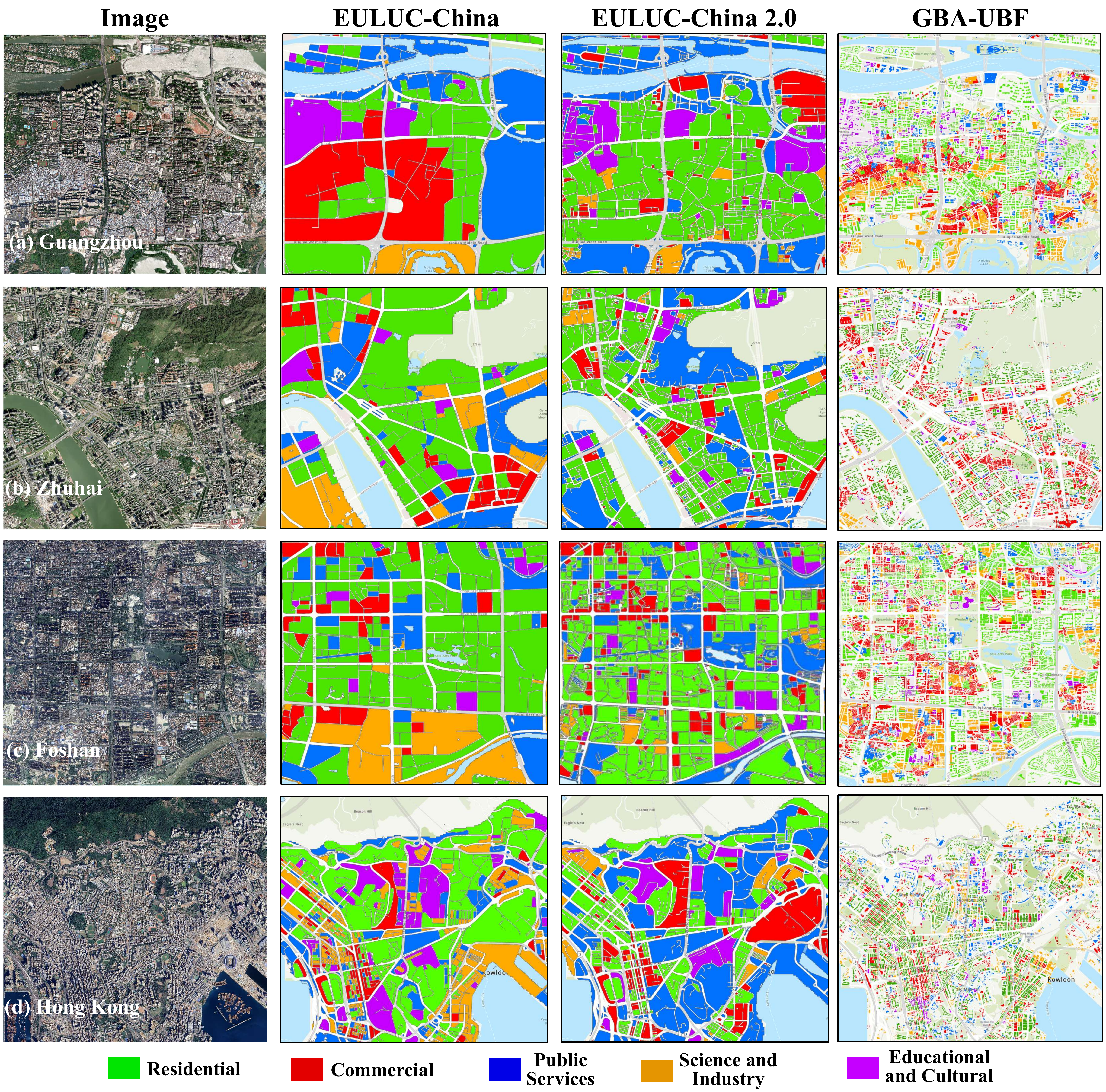}
    \caption{The Comparison between the GBA-UBF Dataset and EULUC-China Dataset.}
    \label{fig:compare}
\end{figure*}

\begin{figure*}[!t]
    \setlength{\abovecaptionskip}{0.1cm}
    \setlength{\belowcaptionskip}{-0.3cm}
    \centering
    \includegraphics[width=0.88\textwidth]{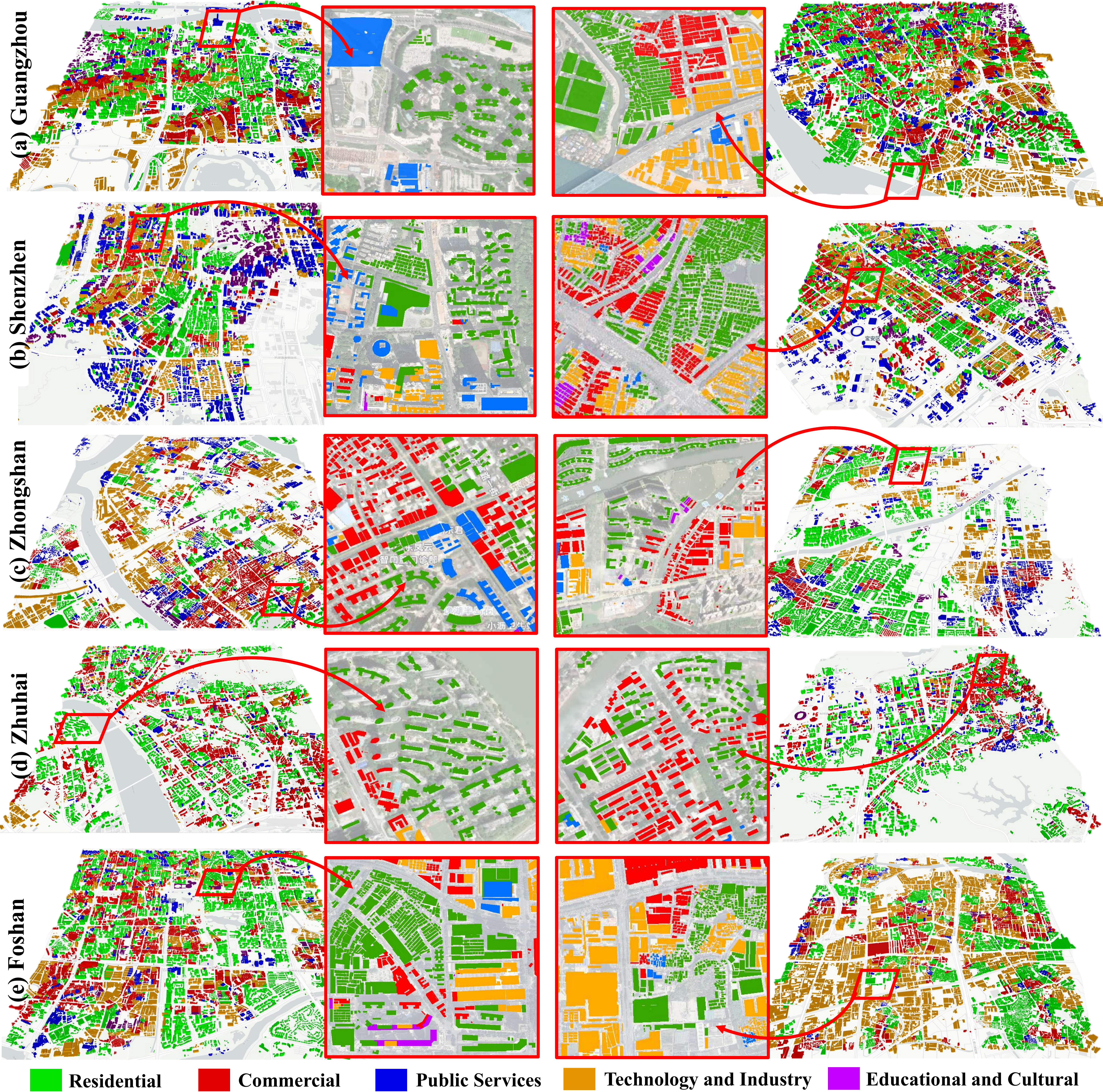}
    \caption{Representative GBA-UBF Dataset 3D Visualization and Corresponding Shapefile Overlay with High-Resolution Remote Sensing Imagery across six cities.}
    \label{fig:3D}
\end{figure*}

\subsection{Qualitative Comparison}

Compared with traditional land use datasets, building function datasets provide finer spatial resolution and higher descriptive accuracy. Conventional land use data are typically organized at the parcel and block level, with coarse granularity and ambiguous boundaries, making it difficult to capture intra-parcel heterogeneity and mixed-use patterns. In contrast, building function datasets are defined at the individual building scale, enabling a more precise representation of functional variations between structures and offering deeper insights into the fine-grained urban space. Moreover, they can integrate multiple data sources (e.g., POI, remote sensing imagery, statistical records), supporting dynamic updates and multidimensional representations. 

The comparison between the \textit{GBA-UBF Dataset} and the \textit{EULUC-China} datasets \cite{li2025enhanced} is shown in Fig.~\ref{fig:compare}. The two EULUC products, while effective at capturing general land-use patterns, operate at the parcel or block level, resulting in coarse boundaries and a limited ability to represent mixed-use structures. Functional heterogeneity within parcels is largely obscured, and many small-scale facilities are generalized into broader classes. In contrast, the GBA-UBF dataset defines functions at the individual building scale, yielding sharper boundaries, finer differentiation of functional types, and greater alignment with the underlying urban morphology. For example, in Guangzhou and Zhuhai, GBA-UBF captures small commercial strips embedded within residential areas, which are omitted in EULUC datasets. Similarly, in Hong Kong and Foshan, GBA-UBF reveals detailed spatial organization of public services, educational facilities, and industrial clusters, offering a more realistic and fine-grained portrayal of urban functions.

Fig.~\ref{fig:3D} shows the 3D view of the representative scenes from six GBA cities. It illustrates precise boundary adherence and clear differentiation among Residential, Commercial, Public Services, Technology \& Industry, and Educational \& Cultural classes. The 3D visualizations reveal vertical morphology, improving label interpretability and supporting downstream analyses and applications.

\begin{figure}[!t]
    \setlength{\abovecaptionskip}{0.1cm}
    \setlength{\belowcaptionskip}{-0.1cm}
    \centering
    \includegraphics[width=0.45\textwidth]{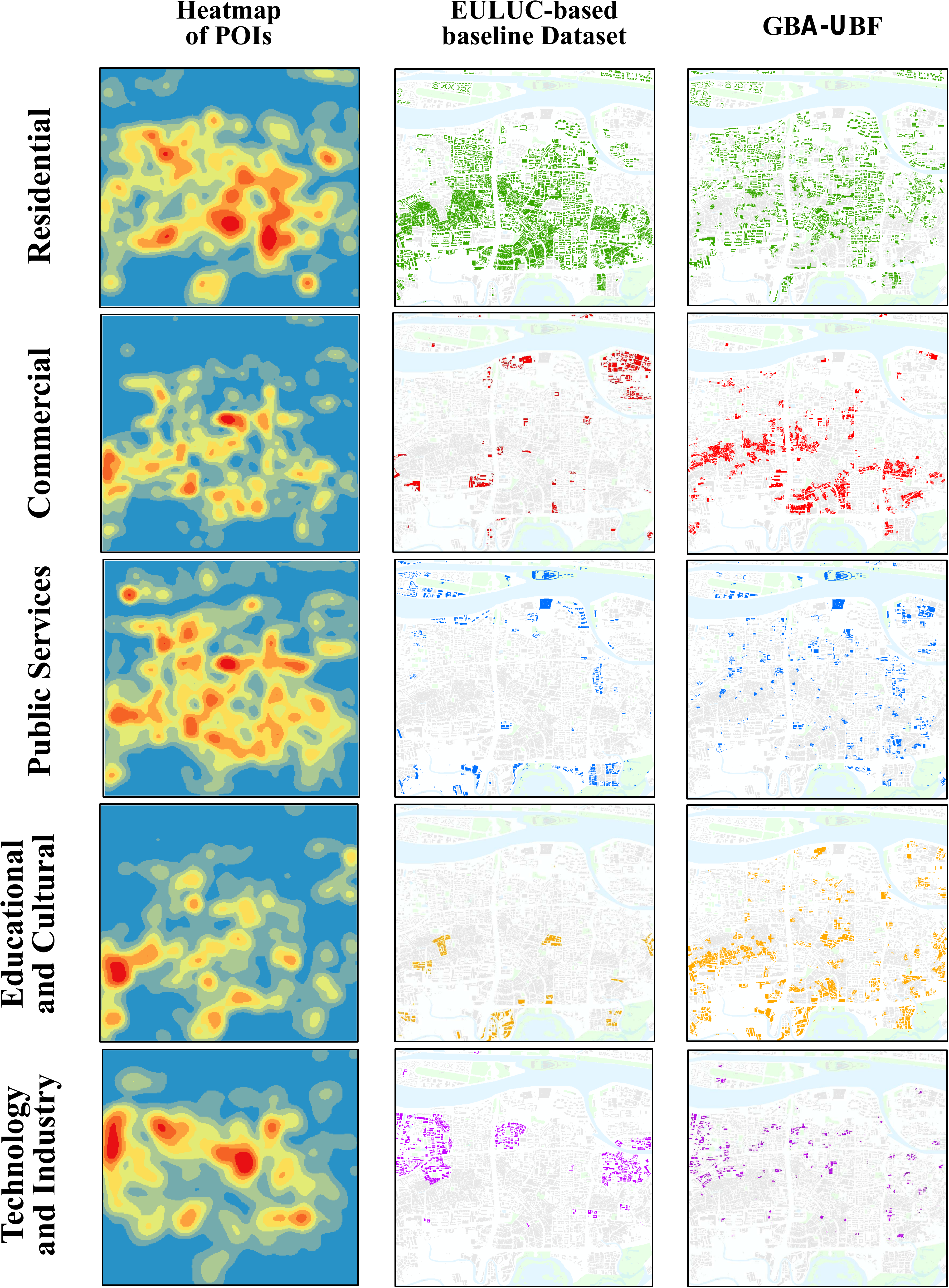}
    \caption{Comparative validation of building function GBA-UBF dataset and land use dataset EULUC-China 2.0 against POI heatmaps.}
    \label{fig:heat}
\end{figure}

We overlay the building footprints with EULUC-China 2.0 \cite{li2025enhanced} function blocks. Each building inherits the function of its corresponding parcel. The results are then consolidated according to the five functional categories defined in this paper, serving as a baseline for comparison. Fig.~\ref{fig:heat} compares the spatial representation of urban building functions. Compared with EULUC-China 2.0, the proposed dataset preserves semantic consistency with POI distributions and provides building-level precision. This demonstrates that our dataset better reflects the true spatial organization of urban functions while maintaining alignment with real-world patterns.

\subsection{Quantitative Comparison}

\begin{figure}[!t]
    \centering
    \setlength{\abovecaptionskip}{0.1cm}
    \setlength{\belowcaptionskip}{-0.5cm}
    \centering
    \includegraphics[width=0.88\columnwidth]{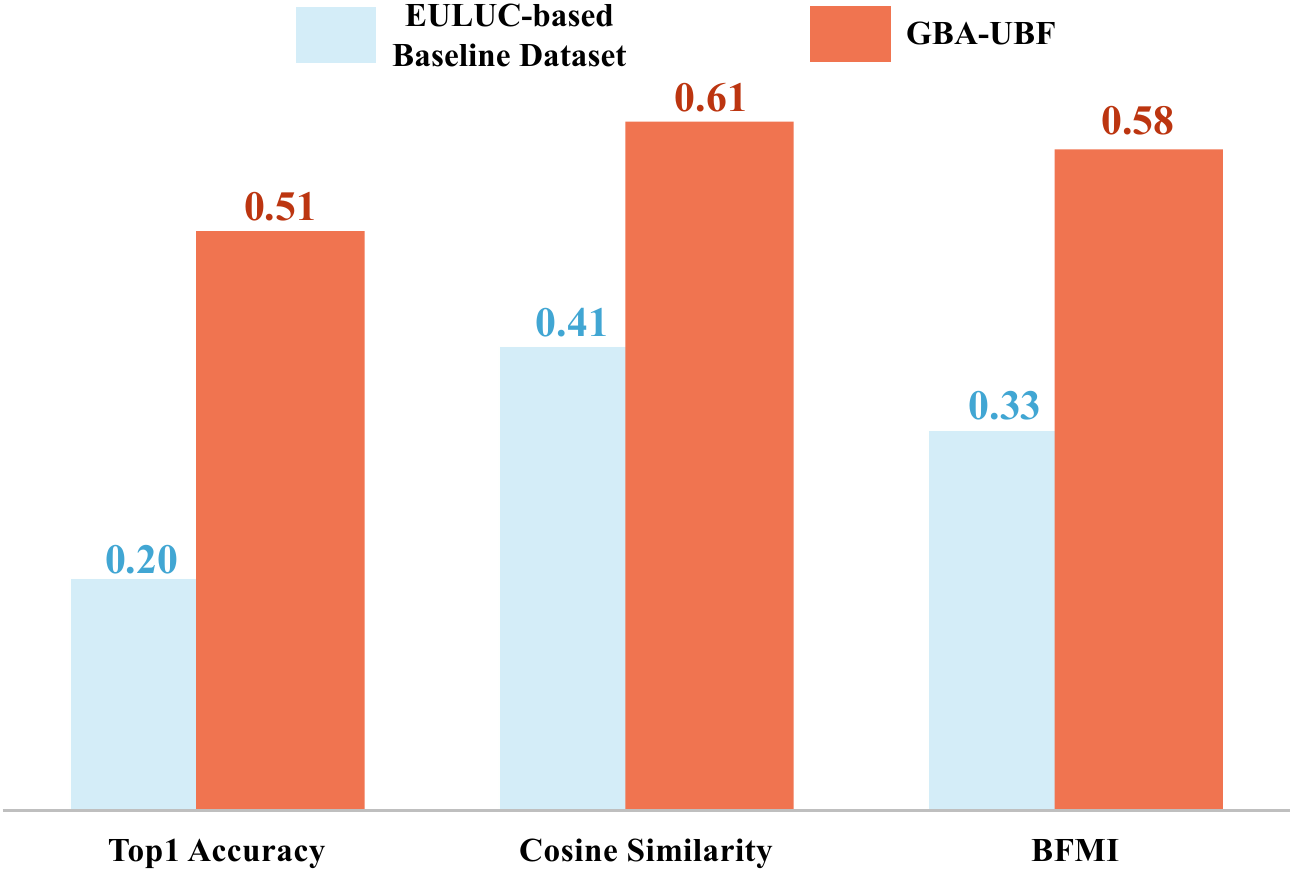}
    \caption{Comparison of Metric Calculations Across Two Datasets.}
    \label{fig:metric}
\end{figure}

The evaluation of quantitative metrics is shown in Fig.~\ref{fig:metric}. Compared with the EULUC-China 2.0, the GBA-UBF demonstrates a consistently superior performance. Specifically, it achieves markedly higher Top-1 Accuracy, indicating that the majority of building labels directly match the reference categories. It also attains substantially higher Cosine Similarity, reflecting better alignment between predicted distributions and POI-derived spatial probabilities. When integrated into the comprehensive BFMI metric, which balances accuracy and distributional similarity, the GBA-UBF dataset outperforms the EULUC-based baseline by a large margin. 

Results indicate that while EULUC-China 2.0 captures dominant building types at the block scale, its granularity remains substantially coarser than our building-level dataset. Because EULUC-China 2.0 is delineated by OpenStreetMap and Tianditu road networks, a single block contains hundreds or even thousands of buildings. As a result, intra-block variation is overlooked, whereas the GBA-UBF dataset provides fine-grained detail at the level of individual buildings. Compared with conventional land use datasets, the building function dataset developed in this study offers broader applicability and superior precision for fine-scale urban research.


\subsection{Field Validation}
To further verify the reliability of the Greater Bay Area Building Function Dataset, we conduct field validation in representative urban sites. This validation aims to cross-check the assigned building functions against ground-truth observations collected from photographic evidence.

Fig.~\ref{fig:overview} illustrates four representative cases. The field validation confirms that the GBA-UBF dataset is capable of distinguishing functionally diverse urban morphologies, even in mixed-use settings. The dataset achieves both semantic interpretability and practical reliability, offering a robust foundation for urban functional zoning and downstream urban analytics.

\subsection{Application Prospects}

Prior studies have demonstrated that functional labeling at the scale of individual buildings significantly enhances the accuracy of urban morphology simulations \cite{BILJECKI2021103440}. In flood risk mapping, heterogeneous exposure datasets—such as residential, industrial, and public service categories—serve as essential inputs for high-resolution damage assessments \cite{FloodRiskBuildingUse}. In the energy sector, the strong coupling between building functions and energy consumption patterns enables block-level estimation of distributed photovoltaic potential \cite{CHENG2020117038}. Similarly, transportation research shows that integrating building functions with travel survey data improves the accuracy of traffic generation rate models by 7–12\% \cite{su13168781}.

From a planning perspective, local governments urgently require high-resolution (1 m) and near-real-time building-level functional datasets when updating regulatory plans or evaluating the transformation of industrial land. Such detailed building-scale data effectively address the limitations of traditional block-level land-use datasets, which lack the granularity needed for contemporary urban management. Consequently, the dataset developed in this study provides not only methodological innovation but also practical value for diverse urban applications.

\subsection{Limitations}

This dataset has several limitations. First, the functional classification adopts a ``single dominant function'' assumption, which overlooks vertical mixed uses, studies in dense cities (e.g., London, New York) show that more than 35\% of buildings host multiple functions \cite{evans2019modelling}. Second, POI timeliness and brand changes introduce label drift, estimated at around 8\% \cite{KONG2024102094}. Third, the dataset excludes attributes such as building age, floors, and structural type, which are critical for applications such as risk assessment and energy modeling \cite{LI2023121217}. Finally, limited POI coverage in Hong Kong and Macau leaves 5–10\% of buildings in newly reclaimed areas without labels, reducing inference confidence.

\section{Conclusion}

This paper presents the Greater Bay Area Urban Building Function Dataset (GBA-UBF), a large-scale dataset that assigns five functional categories to nearly four million buildings across six GBA cities. By integrating building footprints, we design a reproducible three-stage pipeline, achieving fine-grained building-level classification and overcoming the limitations of parcel-based land use data.

To validate quality, we introduce the Building Function Matching Index (BFMI), which combines categorical consistency and distributional similarity against POI-derived heatmaps. Comparative experiments with EULUC-China 2.0 show GBA-UBF delivers markedly higher accuracy and better alignment with urban activity patterns, particularly in dense and mixed-use areas. Field validation further demonstrates its semantic reliability and real-world applicability.

GBA-UBF sets a new benchmark for fine-scale urban analytics, bridging the gap between coarse land-use maps and detailed functional mapping. Its building-level granularity supports diverse applications such as digital twins, flood risk assessment, energy modeling, transport analysis, and policy evaluation, thereby advancing both methodological rigor and practical value in urban geoinformatics and remote sensing.

\section{Acknowledgments}
This work was supported by the National Natural Science Foundation of China under Project 42371343, Guangdong Basic and Applied Basic Research Foundation with Grant No.2024A1515010986.

\bibliographystyle{ACM-Reference-Format}
\bibliography{sample-base}

\appendix







\end{document}